\documentclass[11pt,a4paper]{article}

\usepackage[margin=2.5cm]{geometry} \usepackage{amsmath,amssymb} \usepackage{bbm} \usepackage{xcolor} \usepackage[makeroom]{cancel} \usepackage{microtype} \usepackage{titlesec} \usepackage{fancyhdr} \usepackage{abstract} \usepackage[colorlinks=true, linkcolor=blue!70!black, citecolor=blue!70!black, urlcolor=blue!70!black]{hyperref} \usepackage{anyfontsize}

\usepackage[backend=biber, style=phys, sorting=none, citestyle=numeric-comp, url=false, eprint=false]{biblatex} \appto{\bibsetup}{\sloppy} 

\AtEveryBibitem{%
  \ifentrytype{article}{\clearfield{title}}{}%
  \ifentrytype{inproceedings}{\clearfield{title}}{}%
}

\newcommand{\Tr}{\mathrm{Tr}}

\titleformat{\section}{\large\bfseries}{\thesection.}{0.6em}{}[{\vspace{1pt}\hrule height 0.4pt}\vspace{2pt}] \titleformat{\subsection}{\normalsize\bfseries\itshape}{\thesubsection.}{0.5em}{} \titleformat{\paragraph}[runin]{\normalsize\bfseries}{}{0em}{}[.\enspace]

\newcommand{\affilmark}[1]{\textsuperscript{#1}}

\begin{document}
\thispagestyle{plain}

\begin{center}
  {\fontsize{15}{18}\selectfont\bfseries Unification of Einstein Gravity with Internal Interactions}\\[1.2em] {\normalsize S.\,Stefas\affilmark{$a$}\footnote{E-mail: \href{mailto:dstefas@mail.ntua.gr}{\texttt{dstefas@mail.ntua.gr}}} and G.\,Zoupanos\affilmark{$a,b,c,d$}\footnote{E-mail: \href{mailto:george.zoupanos@cern.ch}{\texttt{george.zoupanos@cern.ch}}} }\\[0.9em] {\small\itshape $^{a}$\,Physics Department, National Technical University of Athens, Zografou Campus, 157\,80 Athens, Greece\\[0.2em] $^{b}$\,Max-Planck Institut f\"ur Physik, Boltzmannstr.\ 8, 85\,748 Garching/Munich, Germany\\[0.2em] $^{c}$\,Universit\"at Hamburg, Luruper Chaussee 149, 22\,761 Hamburg, Germany\\[0.2em] $^{d}$\,Deutsches Elektronen-Synchrotron DESY, Notkestra{\ss}e 85, 22\,607 Hamburg, Germany }
\end{center}

\vspace{0.5em}
\noindent\rule{\linewidth}{0.6pt}

\begin{abstract}
The gauge-theoretic formulation of gravity, enriched with the observation that the tangent group of a curved manifold need not have the same dimension as the manifold itself, provides a natural framework for unifying gravity with internal interactions. In the present work we study the unification of Einstein gravity with the $SO(10)$ grand unified theory within an $SO(1,17)$ gauge theory. As a central result, we derive Einstein gravity with a cosmological constant from the spontaneous symmetry breaking (SSB) of an $SO(1,5)$ gauge theory to the Lorentz group via two independent routes: one using two scalar fields in the fundamental representation $\mathbf{6}$ of $SO(6)$, and one using a single scalar in the adjoint $\mathbf{15}$. Both yield a Gauss-Bonnet + Einstein-Hilbert + cosmological constant action. Embedding the $SO(1,5)$ gravitational sector in $SO(1,17)$ and breaking the full gauge group yields Einstein gravity coupled to $SO(10)$. The Weyl-Majorana condition reduces the fermion family degeneracy to two, in contrast to the four families obtained in the $SO(2,16)$ conformal gravity unification.
\end{abstract}

\section{Introduction}
The unification of all fundamental interactions has been a central aim of theoretical physicists for over a century. One of the well known early attempts was done by Kaluza and Klein \cite{Kaluza:1921,Klein:1926}, who proposed a conceptually new way of unifying gravity and electromagnetism, which were the two well established interactions at that time, namely to extend the spacetime from four to five dimensions. Specifically their proposal was that such a unification could be obtained in five dimensions in a purely gravitational theory, which after dimensional reduction to four dimensions, would lead to a $U(1)$ gauge theory, interpreted as electromagnetism, coupled with gravity. Naturally this approach of unification was considered initially very speculative, it gained renewed interest when it was realized that non-Abelian gauge theories could naturally emerge from similar settings \cite{Kerner:1968, CHO1987358, Cho:1975sf} and could in principle be useful in the description of the Standard Model (SM) of Particle Physics. In particular it was found that a higher-dimensional spacetime of the form $M_D = M_4 \times B$, with $B$ a compact Riemannian manifold with non-Abelian isometry group $S$, was leading, upon dimensional reduction, to a four-dimensional theory describing Einstein gravity coupled to a Yang--Mills gauge theory based on the group $S$, together with scalar fields.

The main success of this construction was that it provided a geometrical explanation of the emergence of gauge theories. Unfortunately though it faced serious problems too, including the lack of a viable classical ground state with the assumed simple direct-product structure and, more notably concerning low-energy physics, that it could not lead to chiral fermions in four dimensions after dimensional reduction \cite{Witten:1983}. These problems could be resolved by introducing Yang–Mills fields in the original action, giving up though the attractive explanation that they emerge geometrically. However in this way higher-dimensional Grand Unified Theories (GUTs) can be easily constructed that include also fermions \cite{Georgi:1974sy, FRITZSCH1975193} with scalars appearing naturally in four dimensions after dimensional reduction as the extra (Lorentz-blind) components of the introduced gauge fields. Concerning the demand to obtain chiral fermions in four dimensions a constraint is imposed on the total dimension of the original theory to be of the form $4n + 2$ \cite{CHAPLINE1982461}. This option gave rise to another unification scheme the Coset Space Dimensional Reduction (CSDR) \cite{forgacs, KAPETANAKIS19924, Kubyshin:1989vd, MANTON1981502}, introduced by Forgacs and Manton (F-M), which could naturally lead to chiral fermions in four dimensions. Inspired by the F-M scheme, Scherk and Schwarz (S-S) developed a very similar reduction using group manifolds \cite{SCHERK197961}, which, although it cannot produce chiral fermions, became very popular in string model building.

It is fair to say that Superstring Theories (SSTs) (see e.g., refs. \cite{Green2012-ul, Lust:1989tj, polchinski_1998}), which were developed a bit later than the CSDR, again with the aim of unifying all fundamental interactions, dominated the investigation of higher-dimensional unification for decades. In addition, they have offered a consistent framework in higher dimensions, with the ten-dimensional heterotic string theory \cite{GROSS1985253}, which stands out as a particularly attractive case. This theory naturally accommodates a GUT based on $E_8 \times E_8$, whose dimensional reduction potentially could lead to the SM. However, it should also be stressed that an experimental confirmation of all higher-dimensional frameworks is still lacking, despite some recent promising developments in the CSDR framework \cite{Chatzistavrakidis:2009mh, Irges:2011de, Manolakos:2020cco, Manousselis_2004, Patellis:2024dfl}.

In our recent works \cite{Roumelioti:2024lvn, Roumelioti:2024jib, Roumelioti:2025cxi, Patellis:2025qbl, Patellis:2025syq, Stefas:2025rcf, Roumelioti:2025cco, Roumelioti:2025nku, Stefas:2025yul} we go further than the above mentioned unification schemes first by treating all interactions, including gravity, as gauge theories and then by keeping the whole discussion directly in four dimensions. We consider it worth recalling here the main historical steps of previous unification attempts for completeness and comparison. In addition though it should be noted that, prior to the revival of interest in higher-dimensional frameworks described above, another significant development was done directly in four dimensions, namely that gravity could be formulated as a gauge theory \cite{weyl, utiyama, kibble1961, Sciama, KAKU1977304, Kaku:1978nz, Ivanov:1980tw, Ivanov:1981wn}. This development together with the well-known fact that the Standard Model of Particle Physics is based on gauge theories paved the way for a natural connection between  gravity and gauge theories, thus pointing towards the unification of all interactions that we follow here. Developments in supergravity \cite{freedman_vanproeyen_2012, Ortín_2015}, which is also based on gauge principles, further pointed in this direction, and the framework has been extended to Noncommutative (NC) gravity \cite{castellani, Chatzistavrakidis_2018, manolakosphd, Manolakos:2022universe, Manolakos:2023hif, Manolakos_paper1, Manolakos_paper2, roumelioti2407}.

The formulation of gravity as a gauge theory started with Weyl \cite{weyl1929, weyl} who related electromagnetism to local phase transformations of the electron field and introduced the vierbein formalism, which became crucial in the gauge formulations of gravity. Then the next important step was done by Utiyama \cite{utiyama} showing that gravity could be treated as a gauge theory of the Lorentz group $SO(1,3)$. The ad hoc introduction of the vierbein in this work was addressed later by Kibble \cite{kibble1961} and Sciama \cite{Sciama}, who suggested that instead of the Lorentz the full Poincar\'e group should be gauged. Then Stelle and West \cite{Kibble:1985sn, stellewest} proposed more refined constructions based on the de Sitter $SO(1,4)$ or anti-de Sitter $SO(2,3)$ groups, which have the same number of generators as the Poincar\'e group, and suggested that Lorentz invariance could be recovered by spontaneous symmetry breaking (SSB). Then the conformal group $SO(2,4)$ was treated as a gauge theory leading to Conformal Gravity (CG) \cite{KAKU1977304} and Fuzzy Gravity (FG) \cite{Chatzistavrakidis_2018, manolakosphd, Manolakos:2022universe, Manolakos:2023hif, Manolakos_paper1, Manolakos_paper2, roumelioti2407} and in turn by SSB to Weyl (WG)\footnote{WG, along with CG is a promising framework for the description of gravity at high energy scales - recent progress in this direction can be found in \cite{Maldacena:2011mk, mannheim, Anastasiou:2016jix, ghilencea2023, Hell:2023rbf, Condeescu:2023izl}.} and Einstein (EG) gravities \cite{Roumelioti:2024lvn}. Moreover the gauging of the supersymmetric extension of the conformal group, the superconformal group, was used in the formulation of $N = 1$ supergravity \cite{freedman_vanproeyen_2012, KAKU1977304}.

Eventually a more direct and ambitious unification proposal within the gauge-theoretic framework has been revived recently \cite{Chamseddine2010, Chamseddine2016, Konitopoulos:2023wst, Krasnov:2017epi, Manolakos:2023hif, Nesti_2010, Nesti_2008, Patellis:2025qbl, Patellis:2024znm, Roumelioti:2025cxi, Roumelioti:2025cco, Roumelioti:2024lvn, noncomtomos}, inspired to a certain extent by earlier attempts \cite{Percacci:1984ai, Percacci_1991, Weinberg:1984ke}. It is based on the observation that the dimensions of the tangent group of a curved spacetime do not necessarily coincide with the dimensions of the manifold itself. Then by employing higher-dimensional tangent groups on four-dimensional spacetime, we could unify gravities and internal interactions in a gauge-theoretic manner. It turned out that a few methods developed originally in the framework of higher-dimensional theories, such as the CSDR \cite{CHAPLINE1982461, Chatzistavrakidis:2009mh, forgacs, Irges:2011de, KAPETANAKIS19924, Kubyshin:1989vd, LUST1985309, Manolakos:2020cco, Manousselis_2004, MANTON1981502, Patellis:2024dfl, SCHERK197961}, can also be applied to this four-dimensional formulation. Nevertheless, challenges such as imposing simultaneously Weyl and Majorana conditions to obtain realistic chiral spectra reappear in this formulation \cite{CHAPLINE1982461, KAPETANAKIS19924}. So far within the gauge theoretic framework a unified theory has been constructed, based on the $SO(2,16)$ which brings together conformal gravity and internal interactions, based on the $SO(10)$ GUT \cite{Konitopoulos:2023wst, Manolakos:2023hif, Patellis:2025qbl, Patellis:2024znm, Roumelioti:2025cxi, Roumelioti:2025cco, Roumelioti:2024lvn}, and has been further extended to unify in a similar way noncommutative (fuzzy) gravity with internal interactions \cite{roumelioti2407}.

In the present work we present the minimal model which unifies Einstein gravity with $SO(10)$.

\section{Einstein Gravity in three dimensions as a gauge theory}
\label{sec2}

In the present section, following ref \cite{Witten:1988hc}, we present the 3-d EG with the aim to establish that it can indeed be described as a gauge theory, as it was advertised in the Introduction and in particular as a Chern-Simons one. In order to demonstrate this important property the appropriate ingredients are the Palatini action, the vielbein and the spin connection in their 3-d version, recalling that in this case the latter two constitute the dynamical variables instead of the metric tensor.

In the vielbein formalism the Einstein-Hilbert (E-H) action without cosmological constant and matter for a 3-d manifold M takes the form:
\begin{equation}
  \label{3dEH}
  S_{EH_3}=\frac{1}{16\pi G}\int_M \epsilon^{
  \mu \nu \rho}e_{\mu}{}^{a} \left(\partial_{\nu} \omega_{\rho a} - \partial_{\rho} \omega_{\nu a} +\epsilon_{a b c} \omega_{\nu}{}^{ b}\omega_{\rho}{}^{c}\right).
\end{equation}
Varying the action with respect to $\omega$ gives the torsionless condition,
\begin{equation}
  T_{\mu \nu}^a=\partial_\mu e_\nu{}^a-\partial_\nu e_\mu{}^a+\epsilon^{a b c} \omega_{\mu b} e_{\nu c}-\epsilon^{a b c} \omega_{\nu b} e_{\mu c}=D_\mu e_\nu{}^a-D_\nu e_\mu{}^a=0
\end{equation}
with
\begin{equation}
  \label{deriv3}
  D_\mu e_\nu{}^a=\partial_\mu e_\nu{}^a + \epsilon^{a b c} \omega_{\mu b} e_{\nu c}.
\end{equation}
Moreover by varying the action wrt $e$ we obtain the vacuum Einstein equations of motion,
\begin{equation}
  R_{\mu \nu a}=\partial_\nu \omega_{\rho}{}^{a}-\partial_\rho \omega_{\nu}{}^{a}+\epsilon_{a b c} \omega_\nu{}^b \omega_\rho{}^c=0\, .
\end{equation}
Above is taken into account that in the 3-d the redefinition $\omega_{\mu}{}^a = \frac{1}{2} \epsilon^{abc}\omega_{\mu bc}$ holds.

Denoting the vielbein and spin connection collectively as a gauge field $A$, we can write the action as $AdA + A^3$, which we recognise as the 3-d general form of a Chern-Simons functional. In turn this observation points towards a possible relation of the 3-d gravity with the Chern-Simons gauge theory. What remains is to find the appropriate gauge group and the corresponding Chern-Simons action and verify that it coincides with the 3-d E-H action, \eqref{3dEH}.

An obvious guess is to consider the 3-d Poincaré group, $ISO(1, 2)$ as the appropriate gauge group. However the Chern-Simons functional is defined on simple Lie groups and therefore is not straightforward to develop a Chern-Simons gauge theory based on this group. What is appropriate is to find an invariant quadratic form on the $ISO(1,2)$ Lie algebra. For the $ISO(1,2)$ group there exists the following invariant non-degenerate quadratic form,
\begin{equation}
  tr(J_a P_b ) = \delta_{ab}\ ,\quad tr(P_a P_b ) = 0\ ,\quad tr(J_a J_b ) = 0\ ,
\end{equation}
with $J_a = \frac{1}{2} \epsilon_{abc}J^{bc}$ the three Lorentz generators and $P_a$ the three translations, which together accommodate the six generators of the $ISO(1,2)$ group and satisfy the following commutation relations,
\begin{equation}
  \label{algebra3}
  \left[J_a, J_b\right] = \epsilon_{abc} J^c\ ,\quad \left[J_a , P_b \right] = \epsilon_{abc} P^c\ ,\quad \left[P_a , P_b\right] = 0 .
\end{equation}
In turn we can write the gauge covariant derivative,
\begin{equation}
  \label{covdev3}
  \tilde{D}_\mu=\partial_\mu+\left[A_\mu, \cdot \right],
\end{equation}
with $A_\mu (x)$ is the gauge connection expanded on the generators of $ISO(1,2)$,
\begin{equation}
  \label{gaugeconnection3}
  A_\mu (x) = e_\mu{}^a(x) P_a + \omega_\mu{}^a(x) J_a\ ,
\end{equation}
i.e. we assign a component gauge field to each generator; the vielbein corresponding to the local translations and the spin connection to the Lorentz transformations.

The $\tilde{D}_\mu$ in eq \eqref{covdev3} transforms covariantly providing the transformation rule $of A_\mu$,
\begin{equation}
  \label{gftransform3}
  \delta A_\mu=-\tilde{D}_{\mu}\epsilon=-\partial_\mu\epsilon-\left[A_\mu,\epsilon\right]\ ,
\end{equation}
with $\epsilon=\epsilon(x)$ the gauge transformation parameter, which, being an element of the $ISO(1, 2)$ algebra, can be expanded on its generators,
\begin{equation}
  \label{infparameter3}
  \epsilon(x)=\xi^a(x)P_a+\lambda^a(x) J_a\ ,
\end{equation}
where $\xi^a(x)$ and $\lambda^a(x)$ are infinitesimal parameters. From eqs \eqref{covdev3}, \eqref{gftransform3} and \eqref{infparameter3} and using the algebra of generators \eqref{algebra3} we find the following transformations of the fields $e$ and $\omega$,
\begin{align}
  \label{deltae}
  \delta e_\mu{}^a & =-\partial_\mu \xi^a-\epsilon^{a b c} e_{\mu b} \lambda_c-\epsilon^{a b c} \omega_{\mu b} \xi_c\, , \\
  \label{deltaomega}
  \delta \omega_\mu{}^a & =-\partial_\mu \lambda^a-\epsilon^{a b c} \omega_{\mu b} \lambda_c \ .
\end{align}
Next we determine the tensors of the gauge fields using the commutator of the covariant derivative of the gauge theory, $D_\mu$,
\begin{equation}
  \label{Rdef3}
  R_{\mu \nu}=\left[\tilde{D}_\mu, \tilde{D}_\nu\right]=\partial_\mu A_\nu-\partial_\nu A_\mu+\left[A_\mu, A_\nu\right]\ ,
\end{equation}
where $A_\mu$ is the gauge connection \eqref{gaugeconnection3}. Given that $R_{\mu\nu}$ is valued in the algebra of $ISO(1, 2)$, we can also write its expansion on the generators of the algebra,
\begin{equation}
  \label{Rxpansion3}
  R_{\mu \nu}=T_{\mu \nu}{}^a(x) P_a+R_{\mu \nu}{}^a(x) J_a\ .
\end{equation}
From eqs.~\eqref{gaugeconnection3}, \eqref{Rdef3} and \eqref{Rxpansion3} we obtain the component curvature tensors,
\begin{align}
  \label{torsion3}
  & T_{\mu \nu}{}^a=\partial_\mu e_\nu{}^a-\partial_\nu e_\mu{}^a+\epsilon^{a b c} \omega_{\mu b} e_{\nu c}-\epsilon^{a b c} \omega_{\nu b} e_{\mu c}\, , \\
  \label{curvature3}
  & R_{\mu \nu}{}^a=\partial_\mu \omega_{\nu a}-\partial_\nu \omega_{\mu a}+\epsilon_{a b c} \omega_\mu{}^b \omega_\nu{}^c\, ,
\end{align}
which are torsion and curvature 2-forms in 3-d.

Finally the gauge theory in 3-d is Chern-Simons action functional,
\begin{equation}
  S_{\mathrm{CS}}=\int_M \operatorname{tr}(A \wedge d A+A \wedge A \wedge A)=\int_M \operatorname{tr} A_\mu\left(\partial_\nu A_\rho-\partial_\rho A_\nu+\left[A_\nu, A_\rho\right]\right) \epsilon^{\mu \nu \rho} d^3 x\ ,
\end{equation}
which substituting the expression for $A_\mu$ from \eqref{gaugeconnection3}, becomes,
\begin{equation}
  \int_M \epsilon^{\mu \nu \rho} e_{\mu}{}^{a} \left((\partial_{\mu} \omega_{\rho a} - \partial_{\rho} \omega_{\nu a} + \omega_{\nu}{}^{b} \omega_{\rho}{}^{c} \epsilon_{abc}) + (\partial_{\nu} e_{\rho a} - \partial_{\rho} e_{\nu a} + (\omega_{\nu}{}^{b} e_{\rho}{}^{c} - e_{\nu}{}^{b} \omega_{\rho}{}^{c}) \epsilon_{abc})
  \right)\ ,
\end{equation}
which means that the action is expressed in terms of the torsion and curvature two-forms given in eqs \eqref{torsion3} and \eqref{curvature3} respectively. The full CS action, now expressed as the sum of these two terms, is invariant under the entire $ISO(1,2)$ gauge group, which includes a local translation symmetry  (generated by $P_a$, parametrized by $\xi^a$) that has no counterpart in standard gravity. Imposing the torsionless condition $T_{\mu\nu}{}^a=0$ eliminates the $\int e \cdot T$ term and constrains the spin connection in terms of the vielbein, reducing the gauge symmetry to the physically appropriate local Lorentz $SO(1,2)$. The action then becomes,
\begin{equation}
  \label{3dCSaction}
  S_{\text{CS}} = \int_M \epsilon^{\mu \nu \rho} e_{\mu}{}^{a}
  \left( \partial_{\nu} \omega_{\rho a} - \partial_{\rho} \omega_{\nu a} + \omega_{\nu}{}^{b} \omega_{\rho}{}^{c} \epsilon_{abc} \right)\ ,
\end{equation}
which, up to a constant, coincides with the E-H action in 3-d.

Finally, it has been shown that the gauge transformations are equivalent to the diffeomorphism transformations \cite{Witten:1988hc}. In other words the invariance of the action under the gauge transformations ensures the general covariance of the theory.

\section{Gauge theory of Einstein Gravity in four dimensions}
\label{sec3}

The way to describe the 4-d gravity as a gauge theory is less straightforward than in 3-d, discussed in Section \ref{sec2}. Let us denote again the vierbein and the spin connection as a gauge field $A$. In four dimensions, following the same procedure of Section \ref{sec2}, the expression of the E-H action would have the form $\sim A\wedge A\wedge(dA + A^2)$, which does not correspond to a 4-d gauge theory. Still, one could construct the Ricci scalar from the curvature tensor $R_{\mu\nu}{}^{ab}$ and recover the E-H action \cite{Ramond:1981pw}. However there exists another, more natural way to obtain the E-H action by treating the Lorentz and translational parts in a unified manner. The basic idea is to start with an action having larger gauge symmetry than the Lorentz group and then break it to $SO(1,3)$ by means of an auxiliary scalar field. The Poincar\'e group $ISO(1,3)$ is an obvious candidate, but the distinct behaviour of its translation generators prevents a full simple-group gauge description. The candidate groups that overcome this are the dS group $SO(1,4)$ and the AdS group $SO(2,3)$, both containing the same number of generators as the Poincar\'e group. Such actions have already been constructed \cite{stellewest, Kibble:1985sn, manolakosphd, Roumelioti:2024lvn} and they yield the E-H action with a cosmological constant after SSB via an auxiliary scalar field.

To be more explicit on the above constructions, first, as in the 3-d case, the vierbein formalism has to be employed. In the absence of a cosmological constant, the isometry group of Minkowski spacetime is the Poincar\'e group $ISO(1,3)$, which in principle should be considered as the gauge group, in accordance with the 3-d case. The Poincar\'e algebra contains ten generators, four local translations $P_a$ and six Lorentz transformations $M_{ab}$, satisfying the following commutation relations,
\begin{equation}
  [M_{ab},M_{cd}]=4\eta_{[a[c}M_{d]b]},\quad[P_{a},M_{bc}]=2\eta_{a[b}P_{c]},\quad[P_{a},P_{b}]=0,
\end{equation}
with $\eta_{ab} = \text{diag}(-1, 1, 1, 1)$ the four-dimensional Minkowski metric.\footnote{The bracket notation implies antisymmetricity of the indices inside, i.e.\ $\eta_{a[b}P_{c]} = \frac{1}{2}(\eta_{ab}P_c - \eta_{ac}P_b)$.} The gauge connection $A_\mu(x)$ expanded on the generators of $ISO(1,3)$ gives
\begin{equation}
  \label{2.65}
  A_\mu(x)=e_\mu{}^a(x)P_a+\frac{1}{2}\omega_\mu{}^{ab}(x)M_{ab}\,,
\end{equation}
with $e_\mu{}^a$ and $\omega_\mu{}^{ab}$ the component gauge fields for translations and Lorentz transformations respectively. The gauge-covariant transformation law for $A_\mu$ is:
\begin{equation}
  \label{2.66}
  \delta A_\mu=D_\mu\epsilon=\partial_\mu\epsilon+[A_\mu,\epsilon]\,,
\end{equation}
with $\epsilon = \epsilon(x)$ a gauge transformation parameter expanded on the generators of $ISO(1,3)$,
\begin{equation}
  \label{2.67}
  \epsilon(x)=\xi^a(x)P_a+\frac{1}{2}\lambda^{ab}(x)M_{ab}\,,
\end{equation}
where $\xi^a(x)$ and $\lambda^{ab}(x)$ are infinitesimal parameters. Combining eqs \eqref{2.65}, \eqref{2.66} and \eqref{2.67} we obtain the expression of the transformation of the component gauge fields,
\begin{align}
  \delta e_{\mu}{}^{a}&=\partial_{\mu}\xi^{a}+\omega_{\mu}{}^{ab}\xi_{b}-\lambda^a{}_{b}e_{\mu}{}^{b}\,,\\
  \delta\omega_{\mu}{}^{ab}&=\partial_{\mu}\lambda^{ab}+\lambda^{a}{}_c\omega_{\mu}{}^{bc}-\lambda^{b}{}_{c}\omega_{\mu}{}^{ac}\,.
\end{align}
Then we obtain the corresponding field strength tensors, $T_{\mu\nu}{}^a$ and $R_{\mu\nu}{}^{ab}$, of the component fields, $e$ and $\omega$, of the gauge connection $A_\mu$ from the definition of the field strength tensor, $R_{\mu\nu}$,
\begin{equation}
  \label{2.70}
  R_{\mu\nu} = [D_\mu, D_\nu] = \partial_\mu A_\nu - \partial_\nu A_\mu + [A_\mu, A_\nu]\, ,
\end{equation}
after its expansion on the generators of $ISO(1,3)$,
\begin{equation}
  \label{2.71}
  R_{\mu\nu} = T_{\mu\nu}{}^a P_a + \frac{1}{2} R_{\mu\nu}{}^{ab} M_{ab}\, .
\end{equation}
Then from eqs \eqref{2.65}, \eqref{2.70} and \eqref{2.71} we obtain the expressions of the component tensors,
\begin{align}
  T_{\mu\nu}{}^a &= \partial_\mu e_\nu{}^a - \partial_\nu e_\mu{}^a - \omega_\mu{}^{ab} e_{\nu b} + \omega_\nu{}^{ab} e_{\mu b}\, ,\\
  R_{\mu\nu}{}^{ab} &= \partial_\mu \omega_\nu{}^{ab} - \partial_\nu \omega_\mu{}^{ab} - \omega_\mu{}^{ac} \omega_\nu{}^{b}{}_{c} + \omega_\nu{}^{ac} \omega_\mu{}^b{}_c\, ,
\end{align}
and we see that the above expressions coincide with those of the torsion and curvature two-form in the vierbein formalism description of general relativity.

Clearly, so far the construction of the gauge-theoretic version of 4-d gravity is straightforward. However, since the Einstein-Hilbert action is not of the usual Yang-Mills type, the desired action must be invariant under the Lorentz and not under the full Poincar\'e symmetry, and a group must be gauged in which all generators are on equal footing. The two candidate groups are the dS group $SO(1,4)$ and the AdS group $SO(2,3)$, which both contain the same number of generators as the Poincar\'e group. Here we present the $SO(1,4)$ case \cite{stellewest}, while the $SO(2,3)$ case has been presented in refs \cite{stellewest, kibble1961, Roumelioti:2024lvn}. The algebra of the $SO(1,4)$ group is given by,
\begin{equation}
  \left[J_{A B}, J_{C D}\right] = \eta_{BC}J_{AD}+\eta_{AD} J_{BC} -\eta_{AC} J_{BD} -\eta_{BD} J_{AC} ,
\end{equation}
with the metric $\eta_{AB} = \text{diag}(-1, 1, 1, 1, 1)$ and $A, B = 1,\dots, 5$. The gauge connection is $A_\mu = \frac{1}{2} \omega_\mu{}^{AB} J_{AB}$, with $J_{AB}$ the ten generators of $SO(1,4)$. From the definition of the field strength tensor of $A_\mu$ we find,
\begin{equation}
  \label{Fstrength}
  F_{\mu \nu}=\left[D_\mu, D_\nu\right]=\partial_\mu A_\nu-\partial_\nu A_\mu+\left[A_\mu, A_\nu\right],
\end{equation}
and given that $F_{\mu\nu} = \frac{1}{2} F_{\mu\nu}{}^{AB} J_{AB}$, we find that
\begin{equation}
  F_{\mu \nu}{}^{A B}=\partial_\mu \omega_\nu{}^{A B}-\partial_\nu \omega_\mu{}^{A B}+\omega_\mu{}^A{}_C \omega_\nu{}^{C B}-\omega_\nu{}^A{}_C \omega_\mu{}^{C B}.
\end{equation}
The only invariant that is polynomial in the field strength tensor is the topological invariant yielding the Pontryagin index,
\begin{equation}\label{initialaction}
  S\sim \int d^4 x \epsilon^{\mu \nu \rho \sigma} F_{\mu \nu}{}^{A B} F_{\rho \sigma A B},
\end{equation}
where $\epsilon^{\mu\nu\rho\sigma}$ is the Levi-Civita symbol. This integral is, however, a total divergence and also parity violating. Although there exists a non-polynomial Lagrangian \cite{West:1978nd}, it is not practical since it contains square roots of the field strength. A preferable choice, also of second order in the field strength, introduces an auxiliary scalar field $\phi^A$ and a dimensionful parameter $m$ with contractions performed using $\epsilon_{ABCDE}$. The resulting action is parity conserving, is not a total divergence, and takes the form,
\begin{equation}
  \label{actionds}
  S=g\int d^4 x\left(m \phi^E \epsilon_{A B C D E} \frac{1}{4}F_{\mu \nu}{}^{A B} F_{\rho \sigma}{}^{C D}\epsilon^{\mu \nu \rho \sigma} +\lambda\left(\phi^E \phi_E-m^{-2}\right)\right),
\end{equation}
with $g$ a dimensionless coupling, and $\lambda$ a parameter acting as a Lagrange multiplier, imposing the constraint
\begin{equation}
  \phi^E\phi_E=m^{-2}
\end{equation}
Picking a specific gauge for the scalar field in which it becomes,
\begin{equation}
  \phi=\phi_0=(0,0,0,0,m^{-1})
\end{equation}
the $SO(1,4)$ gauge symmetry is spontaneously broken down to the little group of $\phi_0$, which is the Lorentz group $SO(1,3)$.\footnote{This is not the unique polynomial Lagrangian choice. In \cite{Wilczek_1998}, combinations of field strength and covariant derivative, or the latter alone, have also been suggested.}

Then the action \eqref{actionds} reduces to the following expression, in which only the Lorentz symmetry is manifest, being again parity conserving,
\begin{equation}
  S_{SO(1,3)}=\frac{g}{4}\int d^4 x \epsilon^{\mu \nu \rho \sigma} F_{\mu \nu}{}^{a b} F_{\rho \sigma}{}^{c d} \epsilon_{a b c d}.
\end{equation}
Defining the scaled gauge fields $e_\mu{}^a = m^{-1} \omega_\mu{}^{a5}$ and decomposing the field strength as $F_{\mu\nu} = \tfrac{1}{2}F_{\mu\nu}{}^{AB} J_{AB} = \tfrac{1}{2}F_{\mu\nu}{}^{ab} J_{ab} + F_{\mu\nu}{}^{a5} J_{a5}$, we obtain,
\begin{equation}\label{tors}
  F_{\mu \nu}{}^{a 5}=m (\partial_\mu e_\nu{}^a-\partial_\nu e_\mu{}^a-\omega_\mu{}^{a b} e_{\nu b}+\omega_\nu{}^{a b} e_{\mu b}),
\end{equation}
and
\begin{equation}\label{curv}
  F_{\mu \nu}{}^{a b}=(\partial_\mu \omega_\nu{}^{a b}-\partial_\nu \omega_\mu{}^{a b}+\omega_\mu{}^{a}{}_{c} \omega_{\nu}{}^{cb}-\omega_\nu{}^{a}{}_{c} \omega_{\mu}{}^{cb})-m^2\left(e_\mu{}^a e_\nu{}^b-e_\nu{}^a e_\mu{}^b\right).
\end{equation}
The expression in parentheses in eq.~\eqref{tors} is the torsion tensor $T_{\mu\nu}{}^a$, while the first parentheses in eq.~\eqref{curv} contain the curvature two-form $R_{\mu\nu}{}^{ab}$ of the vierbein formalism. Setting $F_{\mu\nu}{}^{a5}=0$ yields the torsion-free condition, which determines the spin connection $\omega$ in terms of the vierbein $e$. Substituting the curvature two-form in the broken action we obtain,
\begin{equation}
  \begin{aligned}
    S_{SO(1,3)} &=\frac{g}{4}\int d^4 x \epsilon^{\mu \nu \rho \sigma} \epsilon_{a b c d}\left[R_{\mu \nu}{}^{a b}-m^2\left(e_\mu{}^a e_\nu{}^b-e_\mu{}^b e_\nu{}^a\right)\right]\\
    &\quad\qquad\qquad\qquad\times\left[R_{\rho \sigma}{}^{c d}-m^2\left(e_\rho{}^c e_\sigma{}^d-e_\rho{}^d e_\sigma{}^c\right)\right] \\
    &=\frac{g}{4}\int d^4 x \epsilon^{\mu \nu \rho \sigma} \epsilon_{a b c d}\left[R_{\mu \nu}{}^{a b}R_{\rho \sigma}{}^{c d}-4m^2 R_{\mu \nu}{}^{a b}e_\rho{}^c e_\sigma{}^d+4m^4e_\mu{}^a e_\nu{}^b e_\rho{}^c e_\sigma{}^d\right].
  \end{aligned}
\end{equation}
From the above expression it is clear that the resulting action consists of three terms of the general form
\begin{equation}
  S_{SO(1,3)}=\frac{g}{4}\int d^4 x \epsilon^{\mu \nu \rho \sigma} \epsilon_{a b c d}\left(\mathcal{L}_{R R}- m^2 \mathcal{L}_{R e e}+ m^4\mathcal{L}_{eeee}\right),
\end{equation}
with $g < 0$. The first term yields the Gauss--Bonnet topological invariant and therefore does not contribute to the field equations. The second term can be identified with the E-H action, as it contains the Ricci scalar curvature, and the third term is a cosmological constant of order $m^4$. The field equation $F_{\mu\nu}{}^{ab}=0$ gives positive curvature, so the maximally symmetric solution is de Sitter:
\begin{equation}
  F_{\mu \nu}{}^{a b}=0 \Rightarrow R_{\mu \nu}{}^{a b}=m^2\left(e_\mu{}^a e_\nu{}^b-e_\nu{}^a e_\mu{}^b\right).
\end{equation}
Similar results are obtained by gauging $SO(2,3)$ \cite{West:1978nd, stellewest, Roumelioti:2024lvn} but the resulting maximally symmetric space is AdS.

In conclusion, to obtain the E-H action as a gauge theory one must employ either the dS or AdS group, starting from a polynomial action in the field strength with an auxiliary scalar field satisfying a constraint and fixed to a specific gauge. In this way the original symmetry is reduced to the Lorentz group.

A last comment concerns the general covariance which is recovered by the relation among the gauge transformations and the diffeomorphisms. Following the same procedure and calculations as in the 3-d case, i.e. taking into account the torsionless condition and the equation of motion of vanishing curvature, general covariance is ensured (see e.g.\ \cite{stellewest, Patellis:2025qbl}).

\section{Einstein gravity from \texorpdfstring{$SO(1,5)$}{SO(1,5)}}
\label{sec4}

In this section we study the spontaneous symmetry breaking (SSB) of an $SO(1,5)$ gauge theory to the Lorentz group $SO(1,3)$ and show that the resulting theory is Einstein gravity with a cosmological constant. We present two independent routes to the breaking: one using two scalars in the fundamental representation $\mathbf{6}$ (Route~1), and one using a single scalar in the adjoint representation $\mathbf{15}$ of $SU(4)$ (Route~2). Throughout this section we work in Euclidean signature, exploiting the isomorphism
\begin{equation}
\mathfrak{so}(6) \cong \mathfrak{su}(4),
\label{eq:compactiso}
\end{equation}
which allows us to use the well-developed representation theory of the compact group $SU(4)$ in place of the non-compact $SO(1,5)$; the Lie algebras are isomorphic, so all branching rules, generator matrices and trace identities carry over directly. Throughout this section we use $SO(1,5)$ and $SO(6)$, and similarly $SO(1,3)$ and $SO(4)$, interchangeably.

\subsection{SSB via two scalar fields in the fundamental representation}
\label{sec4.1}

The breaking $SO(1,5)\to SO(1,3)$ can be realised by two scalar fields $\phi^E$ and $\chi^F$, each in the fundamental representation $\mathbf{6}$ of $SO(1,5)\cong SO(6)$. The breaking proceeds in two steps via the branching rules
\begin{equation}
  SO(6)\supset SO(5)\,:\quad\mathbf{6} = \mathbf{1} + \mathbf{5},
  \label{eq:6to5}
\end{equation}
\begin{equation}
  SO(5)\supset SU(2)\times SU(2)\,:\quad
  \mathbf{5} = (\mathbf{1},\mathbf{1}) + (\mathbf{2},\mathbf{2}).
  \label{eq:5to4}
\end{equation}
A vev of $\phi^E$ in the $\langle\mathbf{1}\rangle$ component of~\eqref{eq:6to5} breaks $SO(6)$ to $SO(5)$, and a subsequent vev of $\chi^F$ in the $\langle\mathbf{1},\mathbf{1}\rangle$ component of~\eqref{eq:5to4} breaks $SO(5)$ to $SU(2)\times SU(2)\cong SO(4)$.

To identify which generators are broken at each step, it is useful to decompose the representations of $SO(5)$ under the residual $SU(2)\times SU(2)$:
\begin{equation}
  \mathbf{10} = (\mathbf{3},\mathbf{1}) + (\mathbf{1},\mathbf{3})
                + (\mathbf{2},\mathbf{2}),
  \qquad
  \mathbf{5} = (\mathbf{1},\mathbf{1}) + (\mathbf{2},\mathbf{2}).
  \label{eq:SO5decomp}
\end{equation}
After the first SSB ($SO(6)\to SO(5)$), the five broken generators are $P'_a \equiv J_{a6}$ (four generators, with gauge fields $b_\mu{}^a$) and $Q \equiv J_{56}$ (one generator, with gauge field $\tilde{\alpha}_\mu$), while the ten unbroken generators of $SO(5)$ are the Lorentz generators $M_{ab}$ and the generators $P_a \equiv J_{a5}$. After the second SSB ($SO(5)\to SU(2)\times SU(2)$), the four generators $P_a \equiv J_{a5}$, transforming as $(\mathbf{2},\mathbf{2})$ in~\eqref{eq:SO5decomp}, are broken (with gauge fields $e_\mu{}^a$), leaving only the six Lorentz generators $M_{ab}$, transforming as $(\mathbf{3},\mathbf{1})\oplus(\mathbf{1},\mathbf{3})$, unbroken. Note that the decomposition of the $SO(5)$ adjoint~\eqref{eq:SO5decomp} contains no $(\mathbf{1},\mathbf{1})$ singlet, in contrast to the analogous step in the conformal gravity construction~\cite{Roumelioti:2024lvn}, where the dilatation generator $D$ appears as an additional singlet that must be removed by the 6-plet vev. Here $Q\equiv J_{56}$ was already broken in the first SSB and does not belong to the $SO(5)$ generator algebra.

\subsubsection*{Action and gauge fixings}

The action is
\begin{multline}
  S_{SO(1,5)}
  =
  g\int d^4x\,
  \epsilon^{\mu\nu\rho\sigma}\epsilon_{ABCDEF}\,
  \phi^E\chi^F\,m_\phi m_\chi\,
  \tfrac{1}{4}F_{\mu\nu}{}^{AB}F_{\rho\sigma}{}^{CD}
  \\
  +\;
  g\int d^4x\,\bigl[
    \lambda_\phi(\phi^E\phi_E - m_\phi^{-2})
    +
    \lambda_\chi(\chi^F\chi_F - m_\chi^{-2})
  \bigr],
  \label{eq:action6}
\end{multline}
where $\lambda_\phi$ and $\lambda_\chi$ are Lagrange multipliers enforcing $\phi^E\phi_E=m_\phi^{-2}$ and $\chi^F\chi_F=m_\chi^{-2}$, and $m_\phi\geq m_\chi$. The first scalar is gauge-fixed as
\begin{equation}
  \phi^E = \phi_0 = (0,0,0,0,0,\,m_\phi^{-1}),
  \label{eq:phi0}
  \end{equation}
which is spacelike ($m_\phi^2>0$) and breaks $SO(6)$ to $SO(5)$. The five broken gauge fields associated with the generators $P'_a\equiv J_{a6}$ and $Q\equiv J_{56}$ are rescaled as $A_\mu{}^{a6} = m_\phi b_\mu{}^a$ and $A_\mu{}^{56} = m_\phi\tilde\alpha_\mu$. The second scalar is gauge-fixed as
\begin{equation}
    \chi^m = \chi_0 = (0,0,0,0,\,m_\chi^{-1}),
    \label{eq:chi0}
\end{equation}
which is spacelike ($m_\chi^2>0$) and breaks $SO(5)$ to $SU(2)\times SU(2)$. The four broken gauge fields associated with $P_a\equiv J_{a5}$ rescale as $A_\mu{}^{a5} = m_\chi e_\mu{}^a$. The full identification of the surviving gauge fields is therefore
\begin{equation}
    A_\mu{}^{a6} = m_\phi\,b_\mu{}^a,
    \quad
    A_\mu{}^{56} = m_\phi\,\tilde{\alpha}_\mu,
    \quad
    A_\mu{}^{a5} = m_\chi\,e_\mu{}^a,
    \quad
    A_\mu{}^{ab} = \omega_\mu{}^{ab},
    \label{eq:fieldidentification}
\end{equation}
This is the $SO(6)$ basis expression for the $SO(1,5)$ connection, written in terms of the generators $\{M_{ab}, P_a, P'_a, Q\}$:
\begin{equation}
    A_\mu
    =
    \tfrac{1}{2}\,\omega_\mu{}^{ab}\,M_{ab}
    + e_\mu{}^{\,a}\,P_a
    + b_\mu{}^{\,a}\,P'_a
    + \tilde{\alpha}_\mu\,Q,
    \label{eq:Avector}
\end{equation}
which serves as the starting point for the gravitational interpretation throughout both routes.

\subsubsection*{Field strength components}

Substituting~\eqref{eq:fieldidentification} into the $SO(6)$ field strength and expanding, the components in the $SO(6)$ index space are:
\begin{align}
    F_{\mu\nu}{}^{ab}
    &=
    R_{\mu\nu}{}^{ab}
    - m_\chi^2\bigl(e_\mu{}^a e_\nu{}^b - e_\nu{}^a e_\mu{}^b\bigr)
    - m_\phi^2\bigl(b_\mu{}^a b_\nu{}^b - b_\nu{}^a b_\mu{}^b\bigr),
    \label{eq:Fab6}\\
    F_{\mu\nu}{}^{a5}
    &=
    m_\chi\,T_{\mu\nu}{}^{(0)\,a}(e)
    + m_\phi^2\bigl(\tilde{\alpha}_\mu b_\nu{}^a - \tilde{\alpha}_\nu b_\mu{}^a\bigr),
    \label{eq:Fa5}\\
    F_{\mu\nu}{}^{a6}
    &=
    m_\phi\,T_{\mu\nu}{}^{(0)\,a}(b)
    - m_\chi m_\phi\bigl(\tilde{\alpha}_\mu e_\nu{}^a - \tilde{\alpha}_\nu e_\mu{}^a\bigr),
    \label{eq:Fa6}\\
    F_{\mu\nu}{}^{56}
    &=
    m_\phi\bigl[(\partial_\mu\tilde{\alpha}_\nu - \partial_\nu\tilde{\alpha}_\mu)
    - m_\chi\bigl(e_\mu{}^a b_{\nu a} - e_\nu{}^a b_{\mu a}\bigr)\bigr],
    \label{eq:F56}
\end{align}
where $R_{\mu\nu}{}^{ab}$ is the standard vierbein-formalism curvature of $\omega_\mu{}^{ab}$, and $T_{\mu\nu}{}^{(0)\,a}(e)$, $T_{\mu\nu}{}^{(0)\,a}(b)$ denote the respective torsion tensors. Note that in the Lorentz curvature $F^{ab}$ the clean identification produces only $ee$ and $bb$ bilinears with no $eb$ cross terms, and each of $F^{a5}$ and $F^{a6}$ contains the torsion of a single field ($e$ and $b$ respectively), reflecting the direct correspondence $P_a = J_{a5} \leftrightarrow e_\mu{}^a$ and $P'_a = J_{a6} \leftrightarrow b_\mu{}^a$.

\subsubsection*{Constraints and broken action}

The $\epsilon_{ABCDEF}\phi_0^E\chi_0^F$ projection in the action~\eqref{eq:action6} selects only the $F_{\mu\nu}{}^{ab}$ component, so that $F_{\mu\nu}{}^{a5}$, $F_{\mu\nu}{}^{a6}$, and $F_{\mu\nu}{}^{56}$ are absent from the broken action. Their absence is consistent with imposing the following conditions:
\begin{itemize}
    \item Setting $\tilde{\alpha}_\mu=0$ and $F_{\mu\nu}{}^{a5}=0$
          gives the torsion-free condition for $e$:
          \begin{equation}
            T_{\mu\nu}{}^{(0)\,a}(e) = 0,
            \label{eq:torsione}
          \end{equation}
          which determines $\omega_\mu{}^{ab}$ in terms of $e_\mu{}^{\,a}$.
    \item Setting $\tilde{\alpha}_\mu=0$ and $F_{\mu\nu}{}^{a6}=0$
          gives the torsion-free condition for $b$:
          \begin{equation}
            T_{\mu\nu}{}^{(0)\,a}(b) = 0,
            \label{eq:torsionb}
          \end{equation}
          which determines $\omega_\mu{}^{ab}$ in terms of $b_\mu{}^{\,a}$
          (and is consistent with~\eqref{eq:torsione} once $b=\alpha e$).
    \item Setting $F_{\mu\nu}{}^{56}=0$ with $\tilde{\alpha}_\mu=0$
          yields the algebraic constraint
          \begin{equation}
            e_\mu{}^{\,a}\,b_{\nu a} - e_\nu{}^{\,a}\,b_{\mu a} = 0,
            \label{eq:ebconstraint}
          \end{equation}
          which will also emerge from $F_{\mu\nu}{}^{56}=0$ in Route~2.
\end{itemize}
The broken action is therefore
\begin{equation}
    S_{SO(1,3)}
    =
    \frac{g}{4}\int d^4x\;
    \epsilon^{\mu\nu\rho\sigma}\epsilon_{abcd}\,
    F_{\mu\nu}{}^{ab}\,F_{\rho\sigma}{}^{cd},
    \label{eq:MMaction6}
\end{equation}
which has the same form as eq.~\eqref{eq:MMaction} from Route~2, with $F_{\mu\nu}{}^{ab}$ given by~\eqref{eq:Fab6}.

\subsubsection*{Einstein--Hilbert action}

Imposing $b_\mu{}^a = \alpha\,e_\mu{}^a$ into~\eqref{eq:Fab6}, the $bb$ bilinear becomes $\alpha^2$ times the $ee$ bilinear and the two contributions combine:
\begin{equation}
  F_{\mu\nu}{}^{ab}\big|_{b=\alpha e}
  =
  R_{\mu\nu}{}^{(0)\,ab}
  - (m_\chi^2 + m_\phi^2\alpha^2)
   \bigl(e_\mu{}^{\,a}e_\nu{}^{\,b} - e_\nu{}^{\,a}e_\mu{}^{\,b}\bigr).
  \label{eq:Fab6simple}
\end{equation}
No specialisation $m_\phi=m_\chi$ is required: the result is clean for any values of the two mass scales.\footnote{The earlier works~\cite{Roumelioti:2024lvn,Patellis:2025qbl} used a mixed field identification, combining $b_\mu{}^a$ and $e_\mu{}^a$ into $(b\pm e)$ combinations, motivated by the need to construct commuting translation generators $[P_a,P_b]=0$ on the way to the Poincar\'{e} algebra in the conformal gravity context. That construction required the equal-mass condition $m_\phi=m_\chi$ to yield a clean action, thereby relating the two SSB scales. In the present setting, where $SO(1,5)$ is gauged directly for Einstein gravity without passing through the conformal group, no such requirement on the generators arises and the natural identification is the direct one used here. The absence of cross terms in $F_{\mu\nu}{}^{ab}$ and the resulting clean parametrisation by $\kappa^2 = m_\chi^2 + m_\phi^2\alpha^2$ are consequences of this different starting point.} Substituting into~\eqref{eq:MMaction6} and expanding the square using standard epsilon contractions, the broken action becomes
\begin{equation}
    \begin{split}
      S_{SO(1,3)}
      =
      \frac{g}{4}\int d^4x\;
      \epsilon^{\mu\nu\rho\sigma}\epsilon_{abcd}
      \Bigl[
        &R_{\mu\nu}{}^{(0)\,ab}R_{\rho\sigma}{}^{(0)\,cd} \\
        &-16(m_\chi^2+m_\phi^2\alpha^2)\,
          R_{\mu\nu}{}^{(0)\,ab}\,e_\rho{}^{\,c}\,e_\sigma{}^{\,d} \\
        &+64(m_\chi^2+m_\phi^2\alpha^2)^2\,
          e_\mu{}^{\,a}\,e_\nu{}^{\,b}\,e_\rho{}^{\,c}\,e_\sigma{}^{\,d}
      \Bigr].
    \end{split}
    \label{eq:EH6}
\end{equation}
The three terms are the Gauss--Bonnet topological invariant, the Einstein--Hilbert action, and a cosmological constant. The effective coupling of the EH term is proportional to $m_\chi^2 + m_\phi^2\alpha^2 > 0$ for all values of $\alpha$, and the cosmological constant $\Lambda \propto (m_\chi^2+m_\phi^2\alpha^2)^2 > 0$ is always positive. A positive effective Newton constant requires $g < 0$, and the always-positive cosmological constant confirms de Sitter as the maximally symmetric solution for Route~1.

It is worth noting that the two SSB scales $m_\phi$ and $m_\chi$ enter the action only through the single combination $\kappa^2 \equiv m_\chi^2 + m_\phi^2\alpha^2$, which sets both the effective Newton constant and the cosmological constant via
\begin{equation}
    \frac{1}{16\pi G} \propto g\,\kappa^2,
    \qquad
    \Lambda \propto \kappa^4.
    \label{eq:scales}
\end{equation}
Phenomenological matching to the observed values of $G$ and $\Lambda$ fixes $\kappa^2$ but leaves the ratio $m_\phi/m_\chi$ unconstrained within this section: $m_\phi$ and $m_\chi$ are not individually bounded by the gravitational action alone, only their combination $\kappa^2$ is. The ordering $m_\phi \geq m_\chi$ is the only restriction imposed by the two-step SSB procedure.

\subsection{SSB via an adjoint scalar}
\label{sec4.2}

A second, independent route to the breaking $SO(1,5)\to SO(1,3)$ uses a single scalar field in the adjoint representation $\mathbf{15}$ of $SU(4)\cong SO(6)$. Unlike Route~1, which works in the $SO(6)$ basis~\eqref{eq:Avector} with fields $e_\mu{}^a$, $b_\mu{}^a$, $\omega_\mu{}^{ab}$, $\tilde\alpha_\mu$ identified directly as gauge fields, Route~2 employs the $SU(4)$ basis~\eqref{eq:connection}, in which the breaking pattern and the relevant traces become manifest.

\paragraph{Decomposition and generator identification}

The adjoint representation of $SU(4)$ branches under the subgroup $SU(2)\times SU(2)\times U(1)$ as
\begin{equation}
    \mathbf{15}
    =
    \bigl[(\mathbf{3},\mathbf{1})_0\oplus(\mathbf{1},\mathbf{3})_0\bigr]
    \oplus
    (\mathbf{1},\mathbf{1})_0
    \oplus
    (\mathbf{2},\mathbf{2})_{+2}
    \oplus
    (\mathbf{2},\mathbf{2})_{-2}.
    \label{eq:15decomp}
\end{equation}
The six generators in $(\mathbf{3},\mathbf{1})_0\oplus(\mathbf{1},\mathbf{3})_0$ span the Lorentz subalgebra $\mathfrak{so}(4)\cong\mathfrak{su}(2)\oplus\mathfrak{su}(2)$, the Euclidean rotation group. The singlet $(\mathbf{1},\mathbf{1})_0$ corresponds to the $U(1)$ generator $Y$, and the eight generators in $(\mathbf{2},\mathbf{2})_{+2}\oplus(\mathbf{2},\mathbf{2})_{-2}$ span the broken directions, denoted $Q^{a\dot{b}}$ and $\bar{Q}^{a\dot{b}}$ respectively, with $a,\dot{b}=1,2$.

The $SU(4)$ basis for the unbroken $SU(2)\times SU(2)$ generators is
\begin{equation}
    T^i = \tfrac{1}{2}(J^i + iK^i),
    \qquad
    \widetilde{T}^i = \tfrac{1}{2}(J^i - iK^i),
    \qquad i = 1,2,3,
    \label{eq:chiralbasis}
\end{equation}
where $J^i$ and $K^i$ are the generators of spatial rotations and Lorentz boosts. The coset generators $Q^{a\dot{b}}$ are normalised so that $\Tr[Q^{a\dot{b}}\bar{Q}^{c\dot{d}}] = 2\delta_{ac}\delta_{\dot{b}\dot{d}}$, matching the $SU(2)$ normalisation $\Tr[T^iT^j]=2\delta^{ij}$.

In the $4\times 4$ fundamental representation of $SU(4)$ the generators take the explicit form
\begin{equation}
    T^i
    =
    \begin{pmatrix}\tau^i & 0 \\ 0 & 0
    \end{pmatrix},
    \qquad
    \widetilde{T}^i
    =
    \begin{pmatrix}0 & 0 \\ 0 & \tau^i
    \end{pmatrix},
    \qquad
    Y
    =
    \mathrm{diag}(+1,+1,-1,-1),
    \label{eq:matrices}
\end{equation}
where $\tau^i$ are the Pauli matrices, with $[\tau^i,\tau^j]=2i\epsilon^{ijk}\tau^k$ and $\Tr[\tau^i\tau^j]=2\delta^{ij}$. All commutation relations and trace identities used below follow by direct matrix computation from these explicit forms.

\paragraph{Commutation relations}

The unbroken subalgebra satisfies
\begin{align}
    [T^i,T^j] &= 2i\epsilon^{ijk}T^k, \label{eq:TLTL}\\
    [\widetilde{T}^i,\widetilde{T}^j] &= 2i\epsilon^{ijk}\widetilde{T}^k, \label{eq:TRTR}\\
    [T^i,\widetilde{T}^j] &= 0, \label{eq:TLTR}\\
    [T^i,Y] &= [\widetilde{T}^i,Y] = 0, \label{eq:TY}
\end{align}
and the action of the unbroken generators on the broken sector reads
\begin{align}
    [T^i,Q^{a\dot{b}}]
    &= (\tau^i)_{ca}\,Q^{c\dot{b}},
    \label{eq:TLQ}\\
    [\widetilde{T}^i,Q^{a\dot{b}}]
    &= -(\tau^i)_{\dot{b}\dot{d}}\,Q^{a\dot{d}},
    \label{eq:TRQ}\\
    [T^i,\bar{Q}^{a\dot{b}}]
    &= -(\tau^i)_{ac}\,\bar{Q}^{c\dot{b}},
    \label{eq:TLQbar}\\
    [\widetilde{T}^i,\bar{Q}^{a\dot{b}}]
    &= +(\tau^i)_{\dot{d}\dot{b}}\,\bar{Q}^{a\dot{d}},
    \label{eq:TRQbar}\\
    [Y,Q^{a\dot{b}}]
    &= +2\,Q^{a\dot{b}},
    \label{eq:YQ}\\
    [Y,\bar{Q}^{a\dot{b}}]
    &= -2\,\bar{Q}^{a\dot{b}}.
    \label{eq:YQbar}
\end{align}
The broken generators form a symmetric coset,
\begin{equation}
    [Q^{a\dot{b}},Q^{c\dot{d}}] = 0,
    \qquad
    [\bar{Q}^{a\dot{b}},\bar{Q}^{c\dot{d}}] = 0,
    \label{eq:QQ}
\end{equation}
and the mixed commutator is
\begin{equation}
    [Q^{a\dot{b}},\bar{Q}^{c\dot{d}}]
    =
    (\tau^i)_{ca}\,\delta_{\dot{b}\dot{d}}\,T^i
    \;-\;
    (\tau^i)_{\dot{b}\dot{d}}\,\delta_{ac}\,\widetilde{T}^i
    \;+\;
    \delta_{ac}\,\delta_{\dot{b}\dot{d}}\,Y.
    \label{eq:QQbar}
\end{equation}

\subsubsection*{The $SO(6)$ basis and the $SU(4)$ basis}

As established in Route~1, the $SO(1,5)$ connection decomposes in the $SO(6)$ basis as eq.~\eqref{eq:Avector}. In Route~2 we work instead in the $SU(4)$ basis, in which the same connection is written as
\begin{equation}
    A_\mu
    =
    \omega_\mu{}^i\,T^i
    +
    \tilde{\omega}_\mu{}^i\,\widetilde{T}^i
    +
    B_\mu\,Y
    +
    e_{\mu\,a\dot{b}}\,Q^{a\dot{b}}
    +
    b_{\mu\,a\dot{b}}\,\bar{Q}^{a\dot{b}}.
    \label{eq:connection}
\end{equation}
The spin connection in the two bases is related by
  \begin{equation}
    \tfrac{1}{2}\,\omega_\mu{}^{ab}M_{ab}
    =
    \omega_\mu{}^i\,T^i + \tilde{\omega}_\mu{}^i\,\widetilde{T}^i,
    \label{eq:spinconnrel}
\end{equation}
  which, via the 't~Hooft symbols $\eta^i_{ab}$ (self-dual) and $\bar{\eta}^i_{ab}$ (anti-self-dual), identifies $\omega_\mu{}^i$ and $\tilde{\omega}_\mu{}^i$ as the corresponding projections of $\omega_\mu{}^{ab}$. The $SU(4)$ basis makes the SSB pattern manifest; the $SO(6)$ basis is convenient for the final gravitational interpretation. The $U(1)$ generator $Q\equiv Y$ is the same in both descriptions, and $\tilde{\alpha}_\mu\equiv B_\mu$. The broken-sector fields $e_\mu{}^{\,a}, b_\mu{}^{\,a}$ are linear combinations of the $e_{\mu\,a\dot{b}}, b_{\mu\,a\dot{b}}$ via the standard soldering between vector and bispinor indices.

\subsubsection*{Field strengths}

The field strength tensor
\begin{equation}
    F_{\mu\nu}
    =
    \partial_\mu A_\nu - \partial_\nu A_\mu + [A_\mu,A_\nu]
    \label{eq:Fdef}
\end{equation}
  decomposes in the $SU(4)$ basis as
\begin{equation}
    F_{\mu\nu}
    =
    F_{\mu\nu}{}^i\,T^i
    +
    \widetilde{F}_{\mu\nu}{}^i\,\widetilde{T}^i
    +
    G_{\mu\nu}\,Y
    +
    \mathcal{T}_{\mu\nu}{}^{a\dot{b}}\,Q^{a\dot{b}}
    +
    \bar{\mathcal{T}}_{\mu\nu}{}^{a\dot{b}}\,\bar{Q}^{a\dot{b}}.
    \label{eq:Fdecomp}
\end{equation}
Substituting~\eqref{eq:connection} into~\eqref{eq:Fdef} and using the commutation relations~\eqref{eq:TLTL}--\eqref{eq:QQbar}, the components are
\begin{align}
    F_{\mu\nu}{}^i
    &=
    R_{\mu\nu}{}^i
    +
    (\tau^i)_{ca}
    \bigl(
      e_{\mu\,a\dot{b}}\,b_{\nu}{}^{c\dot{b}}
      -
      e_{\nu\,a\dot{b}}\,b_{\mu}{}^{c\dot{b}}
    \bigr),
    \label{eq:FL}\\
    \widetilde{F}_{\mu\nu}{}^i
    &=
    \widetilde{R}_{\mu\nu}{}^i
    -
    (\tau^i)_{\dot{b}\dot{d}}
    \bigl(
      e_{\mu\,a\dot{b}}\,b_{\nu}{}^{a\dot{d}}
      -
      e_{\nu\,a\dot{b}}\,b_{\mu}{}^{a\dot{d}}
    \bigr),
    \label{eq:FR}\\
    G_{\mu\nu}
    &=
    \partial_\mu B_\nu - \partial_\nu B_\mu
    +
    e_{\mu\,a\dot{b}}\,b_{\nu}{}^{a\dot{b}}
    -
    e_{\nu\,a\dot{b}}\,b_{\mu}{}^{a\dot{b}},
    \label{eq:GY}
\end{align}
where $R_{\mu\nu}{}^i$ and $\widetilde{R}_{\mu\nu}{}^i$ are the standard $SU(2)$ curvatures of $\omega_\mu{}^i$ and $\tilde{\omega}_\mu{}^i$. The coset curvatures are
\begin{align}
    \mathcal{T}_{\mu\nu}{}^{a\dot{b}}
    &=
    D_\mu e_\nu{}^{a\dot{b}} - D_\nu e_\mu{}^{a\dot{b}}
    +
    2\bigl(B_\mu e_\nu{}^{a\dot{b}} - B_\nu e_\mu{}^{a\dot{b}}\bigr),
    \label{eq:Te}\\
    \bar{\mathcal{T}}_{\mu\nu}{}^{a\dot{b}}
    &=
    D_\mu b_\nu{}^{a\dot{b}} - D_\nu b_\mu{}^{a\dot{b}}
    -
    2\bigl(B_\mu b_\nu{}^{a\dot{b}} - B_\nu b_\mu{}^{a\dot{b}}\bigr),
    \label{eq:Tb}
\end{align}
with $D_\mu$ the covariant derivative on the unbroken $SU(2)\times SU(2)$ sector.

\subsubsection*{Action}

The parity-conserving action quadratic in $F_{\mu\nu}$, with an auxiliary scalar field $\Phi$ in the adjoint of $SU(4)$, a dimensionful parameter $m$ and a Lagrange multiplier $\lambda$, is
\begin{equation}
    S
    =
    g\int d^4x
    \Bigl[
      \Tr\bigl(
        \epsilon^{\mu\nu\rho\sigma}\,m\,\Phi\,F_{\mu\nu}F_{\rho\sigma}
      \bigr)
      +
      \lambda\,\bigl(\Phi^2 - m^{-2}\,\mathbbm{1}_4\bigr)
    \Bigr],
    \label{eq:action15}
\end{equation}
where the trace is taken over the generators of $SU(4)$. The Lagrange multiplier $\lambda$ imposes through its variation the constraint
\begin{equation}
    \Phi^2 = m^{-2}\,\mathbbm{1}_4.
    \label{eq:Phi_constraint}
\end{equation}
The scalar is expanded on the full generator basis,
\begin{equation}
    \Phi
    =
    \phi^i T^i + \tilde{\phi}^i\widetilde{T}^i + \phi_Y\,Y
    + \phi_{a\dot{b}}\,Q^{a\dot{b}}
    + \bar{\phi}_{a\dot{b}}\,\bar{Q}^{a\dot{b}}.
    \label{eq:Phiexpansion}
\end{equation}

\subsubsection*{Spontaneous symmetry breaking}

We choose the gauge in which $\Phi$ lies entirely along the $U(1)$ generator $Y$, i.e.\ in the $(\mathbf{1},\mathbf{1})_0$ direction of the decomposition~\eqref{eq:15decomp}:
\begin{equation}
  \Phi = \phi_Y\,Y.
  \label{eq:Phigauge}
\end{equation}
Since $Y^2 = \mathbbm{1}_4$, the matrix constraint~\eqref{eq:Phi_constraint} fixes $\phi_Y = m^{-1}$, giving
\begin{equation}
    \Phi_0 = m^{-1}\,Y.
    \label{eq:vev}
\end{equation}

The breaking pattern can be read off directly from the commutation relations~\eqref{eq:TY}, \eqref{eq:YQ}, \eqref{eq:YQbar}: the generators $\{T^i, \widetilde{T}^i, Y\}$ all commute with $Y$ and span the stabilizer subalgebra $\mathfrak{su}(2)\oplus\mathfrak{su}(2)\oplus\mathfrak{u}(1)$, while $\{Q^{a\dot{b}}, \bar{Q}^{a\dot{b}}\}$ do not. The residual symmetry is therefore
\begin{equation}
    SU(4) \;\longrightarrow\; SU(2)\times SU(2)\times U(1),
    \label{eq:breakingpattern}
\end{equation}
with $\omega_\mu{}^i$, $\tilde\omega_\mu{}^i$, $B_\mu$ as the surviving gauge fields and $e_{\mu\,a\dot{b}}$, $b_{\mu\,a\dot{b}}$ as the coset gauge fields retained as gravitational degrees of freedom.

\subsubsection*{Broken action}

Substituting $\Phi_0 = m^{-1}Y$ into~\eqref{eq:action15}, the Lagrange multiplier term vanishes identically since $\Phi_0^2 = m^{-2}Y^2 = m^{-2}\mathbbm{1}_4$ satisfies~\eqref{eq:Phi_constraint} exactly, and the action becomes
\begin{equation}
    S
    =
    g\int d^4x\;
    \Tr\bigl(
      \epsilon^{\mu\nu\rho\sigma}\,
      Y\,F_{\mu\nu}F_{\rho\sigma}
    \bigr).
    \label{eq:Sred}
\end{equation}
Evaluating the traces from the explicit matrix forms~\eqref{eq:matrices}:
\begin{equation}
    \Tr[YT^iT^j] = +2\delta^{ij},
    \qquad
    \Tr[Y\widetilde{T}^i\widetilde{T}^j] = -2\delta^{ij},
    \label{eq:traceTT}
\end{equation}
\begin{equation}
    \Tr[YT^i\widetilde{T}^j] = 0,
    \qquad
    \Tr[Y^3] = 0,
    \qquad
    \Tr[YT^iY] = \Tr[Y\widetilde{T}^iY] = 0.
    \label{eq:traceszero}
\end{equation}
The two non-vanishing identities in~\eqref{eq:traceTT} follow from $\Tr[\tau^i\tau^j]=2\delta^{ij}$ in $2\times 2$ blocks, with the relative sign coming from $Y$ acting as $+\mathbbm{1}_2$ on the first block and $-\mathbbm{1}_2$ on the second. The vanishing of $\Tr[Y^3]$ (since $Y^2 = \mathbbm{1}_4$, so $Y^3=Y$ and $\Tr[Y]=0$) ensures the $U(1)$ curvature $G_{\mu\nu}$ does not contribute to the broken action. The coset sector also generates trace terms
\begin{equation}
    \Tr[YQ^{a\dot{b}}\bar{Q}^{c\dot{d}}]
    = +2\delta_{ac}\delta_{\dot{b}\dot{d}},
    \qquad
    \Tr[Y\bar{Q}^{a\dot{b}}Q^{c\dot{d}}]
    = -2\delta_{ac}\delta_{\dot{b}\dot{d}}.
  \end{equation}
However, when these are contracted with $\epsilon^{\mu\nu\rho\sigma}$ in the action, the two contributions cancel exactly: relabeling $\mu\nu\leftrightarrow\rho\sigma$ in one of the terms and using the symmetry $\epsilon^{\mu\nu\rho\sigma}=\epsilon^{\rho\sigma\mu\nu}$ gives
\begin{equation}
    \epsilon^{\mu\nu\rho\sigma}\!
    \bigl[
      \mathcal{T}_{\mu\nu}{}^{a\dot{b}}\bar{\mathcal{T}}_{\rho\sigma\,a\dot{b}}
      -
      \bar{\mathcal{T}}_{\mu\nu}{}^{a\dot{b}}\mathcal{T}_{\rho\sigma\,a\dot{b}}
    \bigr]
    = 0.
\end{equation}
The coset curvatures therefore drop out of the action by antisymmetry of $\epsilon^{\mu\nu\rho\sigma}$, not by vanishing of the traces themselves.

The broken action thus reads
\begin{equation}
    S
    =
    2g\int d^4x\;
    \epsilon^{\mu\nu\rho\sigma}
    \bigl(
      F_{\mu\nu}{}^i F_{\rho\sigma}^i
      -
      \widetilde{F}_{\mu\nu}{}^i\widetilde{F}_{\rho\sigma}^i
    \bigr),
    \label{eq:brokenactionchiral}
\end{equation}
where, crucially, the $SU(2)\times SU(2)$ curvatures $F_{\mu\nu}{}^i$ and $\widetilde{F}_{\mu\nu}{}^i$ defined in~\eqref{eq:FL}--\eqref{eq:FR} contain the broken-sector bilinears $e \cdot b$. The fields $e_\mu{}^{\,a}$ and $b_\mu{}^{\,a}$ are therefore present in the broken action as gravitational degrees of freedom.

\subsubsection*{Rewriting in Lorentz notation}

Using the 't~Hooft symbols $\eta^i_{ab}$ (self-dual) and $\bar{\eta}^i_{ab}$ (anti-self-dual), normalised so that $\eta^i_{ab}\eta^{j,ab}=4\delta^{ij}$ and $\tfrac{1}{2}\epsilon_{abcd}\eta^{i,cd}=\eta^i_{ab}$, $\tfrac{1}{2}\epsilon_{abcd}\bar{\eta}^{i,cd}=-\bar{\eta}^i_{ab}$, the $SO(6)$-basis field strength $F_{\mu\nu}{}^{ab}$~\eqref{eq:Avector} decomposes as
\begin{equation}
    F^{ab}_{\mu\nu}
    = F^i_{\mu\nu}\,\eta^{i,ab} + \widetilde{F}^i_{\mu\nu}\,\bar{\eta}^{i,ab}.
    \label{eq:Fab_in_chiral}
\end{equation}
A direct calculation then gives
\begin{equation}
    \epsilon^{\mu\nu\rho\sigma}\,\epsilon_{abcd}\,
    F^{ab}_{\mu\nu}F^{cd}_{\rho\sigma}
    =
    8\,\epsilon^{\mu\nu\rho\sigma}
    \bigl(
      F^i_{\mu\nu}F^i_{\rho\sigma} - \widetilde{F}^i_{\mu\nu}\widetilde{F}^i_{\rho\sigma}
    \bigr),
    \label{eq:chiral_to_Lorentz}
\end{equation}
where the relative minus sign comes from $\eta$ being self-dual and $\bar{\eta}$ anti-self-dual, and the factor of $8$ from the $\eta\eta$ contraction (factor of $4$) combined with the duality (factor of $2$). Substituting~\eqref{eq:chiral_to_Lorentz} into~\eqref{eq:brokenactionchiral} gives the broken action in Lorentz notation,
\begin{equation}
    \boxed{
      S_{\mathrm{SO}(1,3)}
      =
      \frac{g}{4}\int d^4x\;
      \epsilon^{\mu\nu\rho\sigma}\,\epsilon_{abcd}\,
      F^{ab}_{\mu\nu}\,F^{cd}_{\rho\sigma}.
    }
    \label{eq:MMaction}
\end{equation}
No coupling redefinition has been performed: the coefficient $g/4$ follows directly from the trace and projector factors above.

\subsubsection*{Imposed constraints and gravitational interpretation}

The broken action~\eqref{eq:MMaction} depends on $\omega_\mu{}^{ab}$, $e_\mu{}^{\,a}$, $b_\mu{}^{\,a}$, but neither on the coset curvatures $\mathcal{T}_{\mu\nu}{}^{a\dot{b}}$, $\bar{\mathcal{T}}_{\mu\nu}{}^{a\dot{b}}$, nor on the $U(1)$ curvature $G_{\mu\nu}$, nor on $\tilde{\alpha}_\mu\equiv B_\mu$. Since the action is independent of these fields, their vanishing is imposed as a consistent gauge-theoretic projection:
\begin{equation}
    \mathcal{T}_{\mu\nu}{}^{a\dot{b}} = 0,
    \qquad
    \bar{\mathcal{T}}_{\mu\nu}{}^{a\dot{b}} = 0,
    \qquad
    G_{\mu\nu} = 0,
    \qquad
    \tilde{\alpha}_\mu = 0.
    \label{eq:constraints}
\end{equation}
These are not Euler--Lagrange equations of the action~\eqref{eq:MMaction}; they are imposed as gauge-theoretic consistency conditions. \footnote{The same logic was applied in Section~\ref{sec3} for the simpler $SO(1,4)$ case, and a detailed discussion in the gauge-theoretic formulation of conformal gravity may be found in~\cite{Roumelioti:2024lvn}.} The conditions are consistent with the action precisely because the corresponding curvatures are absent from it.

In the $SO(6)$ basis, the vanishing of the coset curvatures $\mathcal{T}_{\mu\nu}{}^{a\dot{b}}$, $\bar{\mathcal{T}}_{\mu\nu}{}^{a\dot{b}}$ becomes the standard torsion-free condition relating $\omega_\mu{}^{ab}$ to $e_\mu{}^{\,a}$ and $b_\mu{}^{\,a}$, while the combined condition $G_{\mu\nu}=0$ with $\tilde{\alpha}_\mu=0$ yields the algebraic relation
\begin{equation}
    e_\mu{}^{\,a}\,b_{\nu a} - e_\nu{}^{\,a}\,b_{\mu a} = 0,
    \label{eq:ebconstraintadj}
\end{equation}
which is the same constraint~\eqref{eq:ebconstraint} derived in Route~1 from $F_{\mu\nu}{}^{56}=0$.

The remaining freedom in eq.~\eqref{eq:ebconstraintadj} is fixed by the ansatz
\begin{equation}
    b_\mu{}^{\,a} = \alpha\,e_\mu{}^{\,a},
    \qquad \alpha\in\mathbb{R}.
    \label{eq:beansatz}
\end{equation}
In Route~2, the broken-sector bilinears in the Lorentz curvatures~\eqref{eq:FL}--\eqref{eq:FR} arise only as $e\cdot b$ cross terms, since the symmetric-coset conditions $[Q,Q]=[\bar{Q},\bar{Q}]=0$ eliminate any $e\cdot e$ or $b\cdot b$ contributions. After the rescaling $e_\mu{}^{\,a}\to m\,e_\mu{}^{\,a}$, $b_\mu{}^{\,a}\to m\,b_\mu{}^{\,a}$ that accompanies the SSB (where $m$ is the single mass scale of Route~2, independent of $m_\phi$ and $m_\chi$ of Route~1), and upon substituting~\eqref{eq:beansatz}, the full Lorentz field strength in the $SO(6)$ basis takes the form
\begin{equation}
    F^{ab}_{\mu\nu}\big|_{b=\alpha e}
    =
    R^{(0)\,ab}_{\mu\nu}
    -4m^2\alpha\bigl(e_\mu{}^{\,a}e_\nu{}^{\,b}-e_\nu{}^{\,a}e_\mu{}^{\,b}\bigr),
    \label{eq:Fab_be}
\end{equation}
where $R^{(0)\,ab}_{\mu\nu}$ is the standard vierbein-formalism curvature of $\omega_\mu{}^{ab}$. The broken action~\eqref{eq:MMaction} then reduces to
\begin{equation}
    \begin{split}
      S_{\mathrm{SO}(1,3)}
      = \frac{g}{4}\int d^4x\;
      \epsilon^{\mu\nu\rho\sigma}\epsilon_{abcd}
      \Bigl[
        &R_{\mu\nu}{}^{(0)\,ab}R_{\rho\sigma}{}^{(0)\,cd}
        -
        16m^2\alpha\,R_{\mu\nu}{}^{(0)\,ab}\,e_\rho{}^{\,c}\,e_\sigma{}^{\,d} \\
        &+
        64m^4\alpha^2\,
        e_\mu{}^{\,a} e_\nu{}^{\,b} e_\rho{}^{\,c} e_\sigma{}^{\,d}
      \Bigr].
    \end{split}
    \label{eq:EHaction}
\end{equation}
The three terms in~\eqref{eq:EHaction} have the standard interpretation: the first is the Gauss--Bonnet topological invariant (which does not contribute to the field equations); the second is the Einstein--Hilbert action with effective gravitational coupling proportional to $m^2\alpha$; and the third is a cosmological constant $\Lambda = 48m^2\alpha$. The sign of $\alpha$ controls the geometry of the maximally symmetric solution: $\alpha > 0$ gives $\Lambda > 0$ (de Sitter), while $\alpha < 0$ gives $\Lambda < 0$ (anti-de Sitter). In both cases a positive effective Newton constant requires $g\,m^2\alpha < 0$. This contrasts with Route~1, where the coefficient $(m_\chi^2+m_\phi^2\alpha^2)>0$ regardless of the sign of $\alpha$, so only the de Sitter geometry arises.

Both routes therefore implement the same group-theoretic breaking $SO(1,5)\to SO(1,3)$ and both yield a Gauss--Bonnet + Einstein--Hilbert + cosmological constant action, confirming that SSB of $SO(1,5)$ leads to Einstein gravity regardless of the choice of scalar representation. The link between the two routes is transparent at the level of branching rules: the adjoint $\mathbf{15}$ of $SO(6)$ decomposes under the intermediate $SO(5)$ as
\begin{equation}
    \mathbf{15} = \mathbf{10} + \mathbf{5},
    \label{eq:15to10plus5}
\end{equation}
where the $\mathbf{10}$ is the adjoint of $SO(5)$ (the generators $M_{ab}$ and $P_a$ left unbroken at the first step of Route~1) and the $\mathbf{5}$ is the coset $SO(6)/SO(5)$ (the generators $P'_a$ and $Q$ broken at that step).

\section{Unification of Gravity and Internal interactions based on SO(1,17)}

\label{sec5}
In this section we describe the unification of Einstein gravity, formulated as a gauge theory with gauge group $SO(1,3)$, with the internal interactions of the $SO(10)$ GUT, which subsequently breaks to the Standard Model. As noted in the Introduction, this approach exploits the observation that the dimension of the tangent group of a curved manifold does not necessarily coincide with the dimension of the manifold itself \cite{Weinberg:1984ke}. The key idea is that the gauge group $SO(1,17)$ contains $SO(1,5)$ as a subgroup, which plays exactly the role studied in Section~\ref{sec4}: upon SSB it yields Einstein gravity with a cosmological constant, as derived there. The remaining factor accommodates the $SO(10)$ GUT and its breaking to the Standard Model.

To obtain a chiral spectrum matching the Standard Model, the gauge group must be of the form $SO(4n+2)$~\cite{Konitopoulos:2023wst}. With the further aim of describing the internal interactions with the $SO(10)$ GUT and imposing the Weyl--Majorana conditions on fermions, we are led to $SO(1,17)$ as the minimal unification gauge group.

\subsection{The Gauge and scalar sector and corresponding SSB}

As in Section~\ref{sec4}, we work in Euclidean signature; the implications of the non-compact space are discussed in Section~\ref{sec5.2}. The starting point is $SO(18)$ with fermions in its spinor representation $\mathbf{256}$, on which the Weyl condition is already imposed. The relevant branching rules under the maximal subgroup $SO(6) \times SO(12)$ are:
\begin{equation}\label{so18}
  \begin{aligned}
      SO(18) & \supset SO(6) \times SO(12) & & \\
      \mathbf{18} & =(\mathbf{6}, \mathbf{1})+(\mathbf{1}, \mathbf{12}) & & \text{ vector} \\
      \mathbf{153} & =(\mathbf{15}, \mathbf{1})+(\mathbf{6}, \mathbf{12})+(\mathbf{1}, \mathbf{66}) & & \text{ adjoint} \\
      \mathbf{256} & =(\mathbf{4}, \overline{\mathbf{32}})+(\overline{\mathbf{4}}, \mathbf{32}) & & \text{ spinor} \\
      \mathbf{170} & =(\mathbf{1}, \mathbf{1})+(\mathbf{6}, \mathbf{12})+(\mathbf{20}^{\prime}, \mathbf{1})+(\mathbf{1}, \mathbf{77}) & & \text{ 2nd rank symmetric}
  \end{aligned}
\end{equation}
The SSB $SO(18)\to SO(6) \times SO(12)$ is achieved by a vev in the $\langle\mathbf{1},\mathbf{1}\rangle$ component of a scalar in the $\mathbf{170}$. The $SO(6)\cong SU(4)$ factor is identified with the gravitational sector studied in Section~\ref{sec4}, while $SO(12)$ accommodates the internal interactions. The Majorana condition can further be imposed on the Weyl fermions in the $\mathbf{256}$, yielding an $SO(6) \times SO(12)$ gauge theory with fermions in the $(\overline{\mathbf{4}}, \mathbf{32})$ representation\footnote{details will be given in Section~\ref{sec5.2}}.

$SO(12)$ is further broken to $SO(10) \times U(1)$ or $SO(10) \times U(1)_{\text{global}}$ by a scalar in the $\mathbf{66}$ (contained in the adjoint $\mathbf{153}$ of $SO(18)$) or in the $\mathbf{77}$ (contained in the $\mathbf{170}$ of $SO(18)$) respectively, with branching rules:
\begin{equation}\label{eq:so12branching}
  \begin{aligned}
      SO(12) & \supset SO(10) \times U(1) \\
      \mathbf{32} & =(\overline{\mathbf{16}})(1)+(\mathbf{16})(-1),\\
      \mathbf{66} & =(\mathbf{1})(0)+(\mathbf{10})(2)+(\mathbf{10})(-2)+(\mathbf{45})(0), \\
      \mathbf{77} & =(\mathbf{1})(4)+(\mathbf{1})(0)+(\mathbf{1})(-4)+(\mathbf{10})(2)+(\mathbf{10})(-2)+(\mathbf{54})(0).
  \end{aligned}
\end{equation}
A vev in the $\langle(\mathbf{1})(0)\rangle$ component of the $\mathbf{66}$ yields $SO(10) \times U(1)$, while a vev in the $\langle(\mathbf{1})(4)\rangle$ component of the $\mathbf{77}$ yields $SO(10) \times U(1)_{\text{global}}$.

The breaking of $SO(6)\cong SU(4)$ to $SO(4)\cong SU(2) \times SU(2)$ proceeds in two steps, following Route~1 of Section~\ref{sec4.1}. The relevant branching rules are:
  \begin{equation}
    \begin{aligned}
      SO(6) & \supset SO(5) \\
      \mathbf{4} & = \mathbf{4},\\
      \mathbf{6} & = \mathbf{1} + \mathbf{5}.
    \end{aligned}
\end{equation}
A vev in the $\langle\mathbf{1}\rangle$ component of a $\mathbf{6}$ of $SO(6)$ (contained in the $\mathbf{18}$ of $SO(18)$) breaks $SO(6)$ to $SO(5)$. Subsequently,
\begin{equation}
    \begin{aligned}
      SO(5) & \supset SU(2) \times SU(2) \\
      \mathbf{5} & =(\mathbf{1},\mathbf{1})+(\mathbf{2},\mathbf{2}),\\
      \mathbf{4} & =(\mathbf{2},\mathbf{1})+(\mathbf{1},\mathbf{2}),
    \end{aligned}
\end{equation}
a vev in the $\langle\mathbf{1},\mathbf{1}\rangle$ component of a $\mathbf{5}$ of $SO(5)$ (contained in the $\mathbf{6}$ of $SO(6)$ and in the $\mathbf{18}$ of $SO(18)$) breaks $SO(5)$ to $SU(2) \times SU(2) \cong SO(4)$. The $\mathbf{4}$ decomposes under $SU(2) \times SU(2)$ into the representations appropriate for the two Weyl spinors.

Alternatively, following Route~2 of Section~\ref{sec4.2}, one may break $SO(6)$ to $SU(2) \times SU(2)$ using a scalar in the adjoint $\mathbf{15}$ of $SU(4) \cong SO(6)$, contained in the adjoint $\mathbf{153}$ of $SO(18)$. The branching rules are:
\begin{equation}
  \begin{aligned}
    SO(6) & \supset SU(2) \times SU(2) \times U(1) \\
    \mathbf{4} & =(\mathbf{2},\mathbf{1})(1)+(\mathbf{1},\mathbf{2})(-1),\\
    \mathbf{15} & =(\mathbf{1},\mathbf{1})(0)+(\mathbf{2},\mathbf{2})(2)+(\mathbf{2},\mathbf{2})(-2)+(\mathbf{3},\mathbf{1})(0)+(\mathbf{1},\mathbf{3})(0).
  \end{aligned}
\end{equation}
A vev in the $\langle\mathbf{1},\mathbf{1}\rangle$ direction of the $\mathbf{15}$ yields the known breaking~\cite{Roumelioti:2024lvn} $SO(6)\to SU(2)\times SU(2)\times U(1)$; the residual $U(1)$ is then eliminated as described in Section~\ref{sec4.2}, yielding $SU(2)\times SU(2)$. The $\mathbf{4}$ again decomposes into the representations required for the two Weyl spinors.

The entire breaking pattern carries over to the non-compact algebras: $SO(18)$ is replaced by $SO(1,17) \simeq SO(18)$, and $SO(6)\cong SU(4)$ by $SO(1,5) \simeq SO(6)$, with branching rules and representations unchanged. The SSB of the $SO(1,5)$ factor to $SU(2)\times SU(2)$ via either Route~1 or Route~2 of Section~\ref{sec4} yields the Einstein--Hilbert action with cosmological constant derived there. The full gauge group after all breakings is therefore $SU(2)\times SU(2)\times SO(10)\times[U(1)]$, where the $SU(2)\times SU(2)$ factor is the Lorentz group and $SO(10)\times[U(1)]$ is the internal sector.

\subsection{Fermion Sector}
\label{sec5.2}

We now address the fermionic sector and the conditions for chirality.

A Dirac spinor in $D$ spacetime dimensions has $2^{D/2}$ independent components; imposing either the Weyl or Majorana condition reduces this by half, and imposing both (when allowed) yields a Weyl--Majorana spinor with $2^{(D-2)/2}$ components. The Weyl condition requires even $D$.

The Dirac, Weyl, Majorana, and Weyl--Majorana spinors transform as finite-dimensional non-unitary representations of $SO(1,D-1)$.

For non-compact groups $SO(p,q)$, Weyl--Majorana spinors exist when $p - q \equiv 0 \pmod{8}$. In higher-dimensional theories such as CSDR~\cite{forgacs, KAPETANAKIS19924, Kubyshin:1989vd, MANTON1981502}, this selects $SO(1,9)$ as the minimal tangent group. In the present four-dimensional gauge-theoretic framework the same constraint applies to the enlarged tangent group; requiring in addition that the internal interactions be described by $SO(10)$ selects $SO(1,17)$ as the minimal unification group \cite{Konitopoulos:2023wst}.

For reference, we collect the relevant four-dimensional spinor notation. Under $SO(1,3)\cong SU(2)\times SU(2)$, the left- and right-handed Weyl spinors $\psi_L$ and $\psi_R$ transform as $(\mathbf{2},\mathbf{1})$ and $(\mathbf{1},\mathbf{2})$ respectively, where $\sim$ denotes "transforms as".

A Dirac spinor can be formed by combining the left- and right-handed Weyl components:
\begin{equation}
  \psi \sim \psi_L + \psi_R \sim (\mathbf{2},\mathbf{1}) + (\mathbf{1},\mathbf{2})
\end{equation}
where in the Weyl basis $\psi_L = (\psi_L, 0)^T$ and $\psi_R = (0,\psi_R)^T$ are eigenstates of $\gamma^5$ with eigenvalues $-1$ and $+1$ respectively. The Majorana condition $\psi = C\bar\psi^T$ (with $C$ the charge-conjugation matrix) connects the $(\mathbf{2},\mathbf{1})$ and $(\mathbf{1},\mathbf{2})$ components and is off-diagonal in the Weyl basis. In even dimensions one may choose a Weyl basis in which $\Gamma^{D+1}$ (the product of all $D$-dimensional gamma matrices) is diagonal:
\begin{equation}
  \Gamma^{D+1} \psi_\pm = \pm\psi_\pm.
\end{equation}
The factorisation $\Gamma^{D+1} = \gamma^5 \otimes \gamma^{d+1}$ separates the four-dimensional and extra-dimensional chiralities; a definite eigenvalue of $\Gamma^{D+1}$ does not fix those of $\gamma^5$ and $\gamma^{d+1}$ separately. Starting from the Weyl spinor $\mathbf{256}$ of $SO(18) \simeq SO(1,17)$ with a fixed $\Gamma^{D+1}$ eigenvalue, the $\gamma^5$ Weyl condition can still be imposed after all breakings, while the Majorana condition can be imposed only once.

After all breakings, the fermion spectrum is:
\begin{equation}
  \begin{aligned}
    SU(2) &\times SU(2) \times SO(10) \times [U(1)]\\
    \{[(\mathbf{2}, \mathbf{1}) &+ (\mathbf{1}, \mathbf{2})\} \{(\overline{\mathbf{16}})(1)+(\mathbf{16})(-1)\}\\
      &=\overline{\mathbf{16}}_L (1)+\mathbf{16}_L(-1)+\overline{\mathbf{16}}_R(1)+\mathbf{16}_R(-1)
    \end{aligned}
  \end{equation}
Using $\overline{\mathbf{16}}_R(1)=\mathbf{16}_L(-1)$ and $\overline{\mathbf{16}}_L(1) = \mathbf{16}_R(-1)$, and projecting onto the $\gamma^5 = -1$ eigenvalue, this reduces to:
\begin{equation}
  2 \times \mathbf{16}_L (-1)\, .
\end{equation}
The construction therefore predicts two fermion families from the group-theoretic structure. The brackets around $U(1)$ indicate that the status of this factor depends on the breaking path of $SO(12)$: using a scalar in the $\mathbf{66}$ (eq.~\eqref{eq:so12branching}) leaves $U(1)$ as a residual gauge symmetry, while using a scalar in the $\mathbf{77}$ (same equation) reduces it to a global symmetry $U(1)_{\text{global}}$. The flavour separation will be addressed in future work.

\section{Conclusions and Discussion}
In our recent papers \cite{Roumelioti:2024lvn, Roumelioti:2024jib, Roumelioti:2025cxi, Patellis:2025qbl, Patellis:2025syq, Stefas:2025rcf, Roumelioti:2025cco, Roumelioti:2025nku, Stefas:2025yul, Konitopoulos:2023wst} (see \cite{Patellis:2025syq} for a comprehensive review) we have presented a systematic framework for the unification of gravity with internal interactions. The approach is based on the gauge-theoretic formulation of gravity and on the observation that the dimension of the tangent group of a curved manifold does not need to coincide with the dimension of the manifold itself, allowing both gravitational and internal interactions to originate from a common gauge structure.

A central result of the present work, developed in Section~\ref{sec4}, is the derivation of Einstein gravity from the SSB of an $SO(1,5)$ gauge theory to the Lorentz group $SO(1,3)$. Two independent routes were presented. Route~1 uses two scalar fields in the fundamental representation $\mathbf{6}$ of $SO(6)$ and breaks $SO(1,5)$ in two steps via the intermediate $SO(1,4)$; Route~2 uses a single scalar in the adjoint $\mathbf{15}$. Both routes yield a Gauss--Bonnet + Einstein--Hilbert + cosmological constant action, confirming the robustness of the construction. The algebraic bridge between them is the branching $\mathbf{15} = \mathbf{10} + \mathbf{5}$ under the intermediate $SO(5)$, where the $\mathbf{10}$ and $\mathbf{5}$ correspond precisely to the unbroken and broken generators of the first step in Route~1. Regarding the geometry of the maximally symmetric solution, Route~1 always yields de Sitter space since the effective cosmological constant is proportional to $\kappa^4 = (m_\chi^2 + m_\phi^2\alpha^2)^2 > 0$ for all values of the ansatz parameter $\alpha$, while Route~2 allows both de Sitter ($\alpha > 0$) and anti-de Sitter ($\alpha < 0$) with $\Lambda = 48m^2\alpha$. Finally, using a direct field identification (in contrast to the mixed combinations of earlier conformal gravity works) the two SSB scales $m_\phi$ and $m_\chi$ of Route~1 enter the gravitational action only through the single combination $\kappa^2$, with no constraint relating them individually within Section~\ref{sec4}.

It is worth noting a further consequence of the two routes to Einstein gravity presented in Section~\ref{sec4}. In the $SO(2,16)$ unification studied in~\cite{Patellis:2024znm}, the breaking of the conformal gravity sector contributes negatively to the cosmological constant, while the $SO(18)$ and $SO(12)$ breakings contribute positively. By analogy with~\cite{Patellis:2024znm}, we expect the $SO(18)$ and $SO(12)$ breakings to contribute positively to the cosmological constant in the present $SO(1,17)$ construction as well. In that case, Route~1 (which always gives a positive contribution from the gravity sector) would result in all breaking contributions carrying the same sign, leaving no mechanism to account for the observed near-vanishing value of the cosmological constant. Route~2 with $\alpha < 0$, on the other hand, gives a negative contribution $\Lambda = 48m^2\alpha < 0$ from the gravity sector, which can in principle balance against the positive contributions from the $SO(18)$ and $SO(12)$ breakings, making this the physically preferred branch in the full unification context. The explicit renormalization group analysis of the $SO(1,17)$ case will be performed in a future publication. 

The gravity theories considered in previous works \cite{Manolakos:2023hif, Roumelioti:2024lvn, Roumelioti:2024jib, Roumelioti:2025cxi, Patellis:2025qbl, Patellis:2025syq, Stefas:2025rcf, Roumelioti:2025cco, Roumelioti:2025nku, Stefas:2025yul} are Conformal Gravity (CG) and Noncommutative (Fuzzy) Gravity (FG), both based on gauging the conformal group $SO(2,4)$ (with FG requiring an additional $U(1)$). Their unification with internal interactions in four dimensions was achieved by gauging the higher-dimensional tangent group $SO(2,16)$, which was broken to Einstein gravity coupled to the $SO(10)$ GUT. In the present work we have shown that an analogous but more direct unification can be realised by gauging $SO(1,17)$. A key feature of this construction is that the Majorana condition can be imposed on fermions in addition to the Weyl condition, reducing the degeneracy of the resulting fermion families. As discussed in Section~\ref{sec5}, for non-compact groups $SO(p,q)$ the existence of Weyl--Majorana spinors requires $p - q \equiv 0\pmod{8}$: consequently, $SO(2,16)$ admits only Weyl spinors and yields four degenerate families, while $SO(1,17)$ admits Weyl--Majorana spinors and yields two. The remaining degeneracy is a known problem; experience with fermion mass degeneracies in the $SO(10)$ GUT suggests that extending the scalar sector to include $\mathbf{120}$- and/or $\mathbf{126}$-dimensional representations can lift it (see~\cite{Mohapatra:book}). We plan to address this in a future publication, together with a renormalization group re-analysis of \cite{Patellis:2024znm} incorporating the necessary additional scalars.

A natural question in any unification of spacetime and internal symmetries is how the Coleman--Mandula (CM) theorem \cite{Coleman1967} is evaded. The CM theorem assumes Poincar\'e invariance as one of its hypotheses. In the present and previous unification schemes, the gauge groups of the gravitational sector ($SO(2,4)$, $SO(2,4)\times U(1)$, and $SO(1,5)$) are all extensions of the Poincar\'e group that are ultimately broken to the Lorentz group by SSB. The resulting theories are therefore not Poincar\'e invariant, so the relevant hypothesis of the CM theorem is not satisfied.

In future work we plan to investigate the cosmological implications of the unification schemes based on $SO(2,16)$ and $SO(1,17)$, in particular the possibility of identifying dark matter with the heavy unobservable particles of the spectrum.

\printbibliography

\end{document}